\documentclass[3p,times]{elsarticle}

\usepackage{ecrc}

\volume{doi:10.1016/j.cpc.2017.03.006}

\firstpage{1}

\journalname{Computer Physics Communications}

\runauth{Y. Guo and V. A. Baulin}

\jid{CPC}

\jnltitlelogo{Comp Phys Comm}
\usepackage{amssymb}
\usepackage[figuresright]{rotating}
\usepackage{algorithm}
\usepackage{algpseudocode,setspace}
\usepackage{amsmath}
\usepackage{amsfonts}
\usepackage{color}
\usepackage[outdir=./]{epstopdf}
\usepackage{booktabs}
\usepackage{multirow}

\begin{document}

\begin{frontmatter}

\dochead{Article}

\title{GPU Implementation of the Rosenbluth Generation Method for Static Monte Carlo Simulations}

\author{Yachong Guo$^a$}

\author{Vladimir A. Baulin$^{a,*}$}
\address[Universitat Rovira i Virgili]{Departament
	d'Enginyeria Quimica, Universitat Rovira i Virgili 26 Av. dels
	Paisos Catalans, 43007 Tarragona, Spain}

\begin{abstract}
We present parallel version of Rosenbluth Self-Avoiding Walk generation method implemented on Graphics Processing Units (GPUs) using CUDA libraries. The method scales almost linearly with the number of CUDA cores and the method efficiency has only hardware limitations. The method is introduced in two realizations: on a cubic lattice and in real space.
We find a good agreement  between serial and parallel implementations and consistent results between lattice and real space realizations of the method for linear chain statistics. The developed GPU implementations of Rosenbluth algorithm can be used in Monte Carlo simulations and other computational methods that require large sampling of molecules conformations.
\end{abstract}


\end{frontmatter}

\section{Introduction}

Statistical methods and computer simulations play major role in theoretical understanding of many-body interactions in physics and chemistry \cite{frenkel_understanding_2002}. In particular, Molecular Dynamics (MD) and Monte Carlo (MC) methods \cite{leach_molecular_2001} are the main theoretical tools used to describe physical and chemical processes at the molecular level. Increasing computer power and availability of computational recourses contribute in growing popularity of computational methods. Even computationally expensive \textit{ab-initio} calculations become feasible nowadays: the length-scales and time-scales of atomistic simulations increased more than 10 times in a decade \cite{muller_biological_2006}.

However, mostly used computational methods such as MD and MC simulations and numerous computational techniques were conceived at the beginning of computer era in the late 1950s \cite{alder_studies_1959}, when a rigid architecture of single-core microprocessors imposed on the structure of the theoretical methods in form of a list of instructions for Central Processing Unit (CPU) implemented sequentially. Miniaturization of processors and increase of clock speed is reaching the physical limit \cite{sutter_software_2005} impeding further increase of computational efficiency. To handle that problem computer industry explores two main paths: multi-core and many-thread processors \cite{hwu_concurrency_2008}. Both ways assume parallelization of tasks and synchronous work with data.

Rapidly growing industry of Graphics Processing Units (GPUs) driven by fast-growing video game market provides new dimension in computational resources: a current example is the NVIDIA Tesla K40 that can reach about 5 trillion floating-point operations per second, while a new released Intel Core i7-5960K (Xenon Haswell) processor can only reach 350 billion floating-point operations per second. This makes GPU very attractive for scientific computation\cite{owens_gpu_2008} and many traditional scientific methods including MD, MC, finite element analysis are adapting for GPU. As a result, GPU versions of the codes are accelerated by factors from 10 to 100 compared to single core CPUs\cite{januszewski_accelerating_2010,rapaport_enhanced_2011,nedelcu_gpu_2012,komura_gpu-based_2012,anderson_general_2008,preis_gpu_2009}.

Nevertheless, the adapted parallel versions of traditional methods cannot use full advantage of GPU architecture, because they were designed conceptually for sequential implementation and thus, contain large portions of  non-parallilizible parts of the code and inter-connections that require communication between the cores, for example, to update the list of nearest neighbors. Thus, there is a need in the development of new methods that are specially designed for modern highly parallel architecture. 

In the present work we present a highly parallel version of Rosenbluth method which is in the ground of static MC simulations first introduced in 1955 \cite{rosenbluth_monte_1955}. Two parallel implementations of the method on graphics processors units (GPU) of the Rosenbluth method are presented: on the lattice and in real space leading to drastic speed increase in simulation in polymer and soft matter science.

The paper is organized as follows. After brief description of static MC methods in Section 2, we describe the GPU implementation of Rosenbluth sampling method in Section 3. Comparative examples between serial and parallel  implementation of the Rosenbluth method for polymer chains up to 64 monomers are presented in Section 4. We summarize our results in Section 5.

\section{Static Monte Carlo methods}

In equilibrium statistical mechanics thermodynamic properties are represented by the ensemble averages of the observable $A$ over all coordinates of $N$ particles $r^N$.

\begin{equation}
\left\langle A\right\rangle =\dfrac {\int dr^{N}A(r^{N})\exp\left[-\beta\mathcal{U}\left( r^{N}\right)\right]  } {\int dr^{N}\exp\left[-\beta\mathcal{U}\left( r^{N}\right)\right] }
\end{equation}
where $ \beta = 1/k_BT$, where $\mathcal{U}$ is the potential energy of the system.
In general, the integral cannot be solved analytically, however, MC simulations provide a numerical approach to this problem by generating a random sample of configuration space points $r^N_1,r^N_2...r^N_\Gamma$, due to the high degree of freedom depending on $N$, it is in general not possible to sample the entire original distribution, one can use a similar but smaller sampling distribution $P_s(r^N)$ to replace the original distribution then correcting for the corresponding error. Such technique is known as representative sampling \cite{kruskal_representative_2004}. $\left\langle A\right\rangle$ is then estimated by

\begin{equation}
\overline {A}=\dfrac {\sum\limits _{\gamma=1}^{\Gamma}A\left( r^{N}_\gamma\right) \exp\left[-\beta\mathcal{U}\left( r_\gamma^{N}\right)\right]/P_s(r_\gamma^N) } {\sum\limits _{\gamma=1}^{\Gamma} \exp\left[-\beta\mathcal{U}\left( r_\gamma^{N}\right)\right]/P_s(r_\gamma^N) }
\label{MD}
\end{equation}
Whether $\overline {A}$ represents a good estimate for $\left\langle A\right\rangle$ depends on the total number $\Gamma$ of configurations used and, for a given $\Gamma$, on the choice of $P_s(r^N)$, 
which, in turn, should approximate $\exp\left[-\beta\mathcal{U}\left( r^{N}\right)\right]$ as closely as possible to obtain meaningful results from MC simulations. MC simulations can be static or dynamic. In this paper we mainly focus on the static MC, where a sequence of statistically independent configuration-space points from the distribution $P_s(r^{N})$ is generated as a basic sampling. 

There exists a large class of sampling algorithms based on MC methods. Rosenbluth sampling is a MC method for generating correctly distributed Self Avoiding Walk (SAW) by means of weights calculated on the fly. Rosenbluth sampling method have been widely applied and used due to its efficiency and simplicity of implementation.

The basic idea of Rosenbluth sampling is to avoid self-intersections by only sampling steps leading to self-avoiding configurations. Hence the algorithm will terminate only when the walk is trapped in a dead end and cannot continue growing. Although this still happens exponentially often for long chain, Rosenbluth sampling can produce substantially longer configurations than simple sampling. During Rosenbluth generation process \cite{rosenbluth_monte_1955}, a monomer can be placed to adjacent sites which can be selected with a probability $p$. The weight of the generated configuration is multiplied by $1/p$. Thus, $n$-step walk grown by Rosenbluth sampling has a weight

\begin{equation}
W_{n}=\prod _{i=0}^{n-1}\dfrac {1} {p_{i}}
\label{Roseq}
\end{equation}
where $1/p_{i}$ is the number of ways in which a configuration can continue to grow after $i$-th growth step. This walk is generated with the probability $P_n = 1/W_n$. Eq. \ref{Roseq} shows that configurations with lower $p_{i}$ have a lower probability of occurring. This bias toward dense configurations in the production of a SAW is corrected in calculation of averages by the weight $W$ when calculating observables, see Eq. \ref{MD}. 

\section{GPU Implementation of Rosenbluth algorithm}
\subsection{Space discretization}

For a lattice implementation, we subdivide the simulation box into $M \times M \times M$ (M $\in  \mathbb{N}^+$) lattice units, where each lattice unit can be occupied by only one monomer. The bond length equals the lattice constant, and the bond angles are restricted by the lattice geometry. 

In the case of off-lattice implementation, the chains are represented as a sequence of beads which can be placed randomly in 3D space according to SAW.

\subsection{Random Number Generator}

Generation of a representative sampling conformations of polymers require random numbers with long periods and good statistical properties.  Generating pseudo-random numbers on a CPU is a well-studied topic \cite{matsumoto_mersenne_1998,lecuyer_random_1990,park_random_1988}, in a GPU Single Instruction Multiple Thread (SIMT) environment, many approaches have been used for the generation of random numbers in different types of applications \cite{nickolls_scalable_2008,farber_cuda_2011}. The description of Random Number Generators (RNGs) can be found in the literature for single stream computations \cite{matsumoto_mersenne_1998,lecuyer_random_1990}, or parallel implementations\cite{luscher_portable_1994,oconnor_spss_2000}. Any RNG chosen should guarantee that random numbers to be generated and immediately consumed by user kernels without requiring the random numbers to be written to and then read from global memory. It also guarantees that each thread generates their own random number at the same time. In our code we chose the Mersenne twister \cite{metropolis_equation_1953}, which guarantees uncorrelated random number streams of each thread. A detailed implementation of Mersenne twister on GPU can be found, for example, in the SDK library from NVIDIA \cite{_cuda_????}.

We initialize the RNG with a single random number seed but a different sequence number for every thread. To initialize the Mersenne twister generator \cite{matsumoto_mersenne_1998}, it is necessary to create a RNG \cite{matsumoto_mersenne_1998} status for every thread and pass this status to the $Curand\_init$ \cite{_cuda_????} function with a seed but different sequence number. The distance between the first elements in successive sequence for Mersenne twister is $2^{67}$, so that it is unlikely that two sequences will overlap even in extensive simulation. In our implementation, we initialize the Mersenne twister once and use the updated RNG status for the entire calculation. Once the RNG is initialized, a normally distributed pseudo-random numbers can be generated for all individual threads.
In current implementation, the initial seed was given before the starting of the sampling kernel.

\subsection{Kernel implementation}

\subsubsection{Data structure}

The proper choice of the data structure is critical for implementation performance. In the present work, all the coordinates of each monomer are stored in a shared memory during the SAW process, the coordinates of each conformation and corresponding Rosenbluth weight are flushed to the global memory when the chain is successfully generated.

For lattice implementation, the coordinates of the monomer therefore are stored as numerical integer type, which can benefit from half-precision introduced in new architecture Pascal. For real-space implementation, the coordinates of the monomer are stored as single-precision or double precision floating-point type. However, double precision greatly increases the shared memory consumption leading to push the GPU into the occupancy limitation and increasing bank conflicts, thus decreasing the performance. That is why all calculations are performed using single-precision floating-point operations. We expect that the performance of GPU with double-precision environment will be improved in the future chip generations. The length of the chains in our calculations is limited due to limitations of shared memory in existing GPU chips. The memory architecture and shared memory size may greatly improve in future GPU architectures thus allowing for increased performance for longer chains.

In a parallel perspective, for each block on GPU, $D$ denotes the dimension of the space, $BlockDim.x$ is the block dimension defined by user and $chainlength$ is the total number of conformations to be generated. With this, $D$ arrays with dimensions $BlockDim.x$*$chainlength$ are allocated in the shared memory to store the coordinates. Same numbers of arrays with dimensions $BlockDim.x$ are allocated in the shared memory to store a temporary position of the subunit during molecule generation. The variables, such as Rosenbluth weight, the increment counter of overlaps depend on the thread index, and are stored in the shared memory; each with the size of $BlockDim.x$ to distinguish between different configurations.

\subsubsection{Cubic lattice realization}

\begin{figure}[H]
	\centering
	\includegraphics[width=12cm]{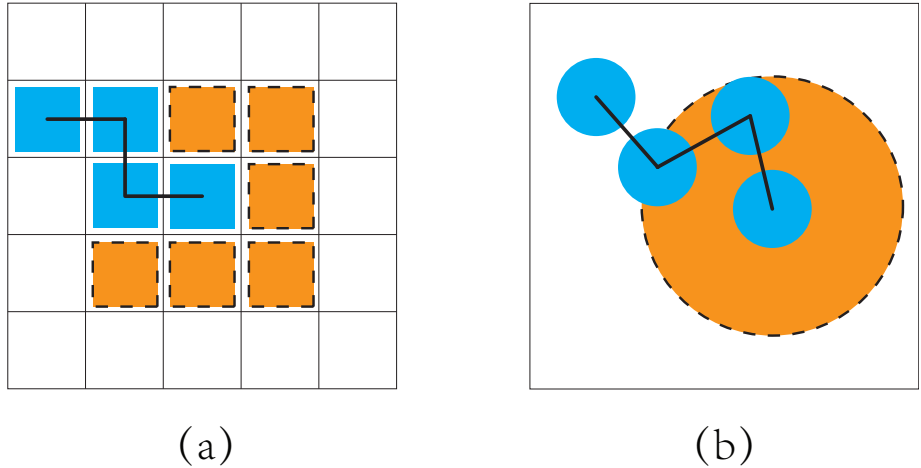}
	\caption{Schematic representation of chain generation in 2 dimension space: (a) on a lattice space (b) in a real space}
	\label{fig:cuda}
\end{figure}

The full computational task of parallel random monomer selection and random monomer placement can be programmed at once in a single GPU kernel. This will give opportunity to use GPU parallelization to maximum extent, limited only by hardware restrictions: GPU memory restrictions on coalesced reads and writes, and more importantly on register use. Each polymer conformation is generated in a separate CUDA thread.

\floatname{algorithm}{Algorithm}
\renewcommand{\algorithmicrequire}{\textbf{Require:}}
\renewcommand{\algorithmicensure}{\textbf{Ensure:}}
\algnewcommand\True{\textbf{True}\space}
\algnewcommand\False{\textbf{False}\space}

\begin{algorithm}
	\setstretch{1.00}
	\caption{Rosenbluth chain generation on a cubic lattice}
	\label{alg:lattice}
	\begin{algorithmic}[1] 		
		\Require $BlockDim.x$ is the block dimension defined by user
		\Require $seed$ is a random number seed chosen by the user
		\Require $chainlength$ is the polymer chain length chosen by the user
		\Require \textbf{Function} $rngonlattice$ generates a random position for the next monomer
		\Require \textbf{Function} $distance$ calculates the distance between two monomers
		\State $x,y,z \gets BlockDim.x * chainlength$
		\State $w \gets BlockDim.x$
		\State $rng \gets rng(seed)$
		\Function {Chaingeneration}{}
%
		
		\For{$i = 0 \to chainlength$}
		\If {$i == 1$}
		\State $Pos[0] \gets 0$ \Comment{$Pos[i]$ denotes position of monomer $i$ $(x_i,y_i,z_i)$}
		
		\State $p \gets 1 /6$\Comment{$p$ denotes Rosenbluth weight}
		\EndIf	
		
		\For{$k = 0 \to ptMax$} \Comment{$ptMax$ denotes total trial attempts}
		\State $Pos[Temp] \gets rngonlattice$
		\State $Pos[Temp]=Pos[i-1]+Pos[Temp]$
		\State $overlap \gets \False$
		
		\For{$j = 0 \to i-1$}
		\If {$Pos[j] \wedge Pos[Temp]$}
		\State $overlap \gets \True$
		\EndIf
		\EndFor
		\If {$\neg overlap $}
		\State $Pos[i]=Pos[Temp]$
		\State $break$
		\EndIf
		\EndFor
		\EndFor
		
		\For{$i = 1 \to chainlength-1$}
		\State $KK \gets 0$
		\For{$j = 0 \to i-1$}
		\State $R2 \gets distance(Pos[i],Pos[j])$
		\If {$R2 \leq d_{max}$} \Comment{$d_{max}$ denotes the maximum polymer bondlength}
		\State $KK=KK+1$
		\EndIf
		\EndFor
		\State $p=p*(1.0/(Cod-kk))$
		\EndFor			
		\EndFunction		
	\end{algorithmic}
\end{algorithm}

The number of blocks $NumBlock$ is defined as the total number of chains to be generated divided by the number of threads per block. In this way each conformation in each block is tagged by the thread number while generated. All the coordinates and weights associated with each conformation are stored in the shared memory. The first monomer can be placed at the center of the coordinate system or it can be placed at any position in the simulation box. The second bond vector is randomly chosen within all $z=26$ possible lattice space in 3D space and added to the monomer. Starting from the third monomer, a bias is introduced to the position of a new monomer due to possibility to overlap with previous monomers. This is taken into consideration in the Rosenbluth weight, Eq. \ref{Roseq} which is by definition the probability of positioning a new monomer without overlaps with previous monomers. In the lattice model, the probability can be easily calculated by looking into the occupancy of previous generated beads in the $z=26$ possible lattice space. Thus, In the lattice model, we can calculate the Rosenbluth weight after the whole chain is generated and thus to reduce the number of calculations. 

The random chain growth process will be continued if the chain is self-avoiding. However, if the chain grow to a dead end when all possible nearby cells are occupied by previous monomers, the whole generated sequence, obtained up to this point, must be discarded, and start at the first step again.
To repeat these steps we can get the SAW of a linear chain of length $N$, see Algorithm \ref{alg:lattice}. 
\begin{figure}[H]
	\centering
	\includegraphics[width=12cm]{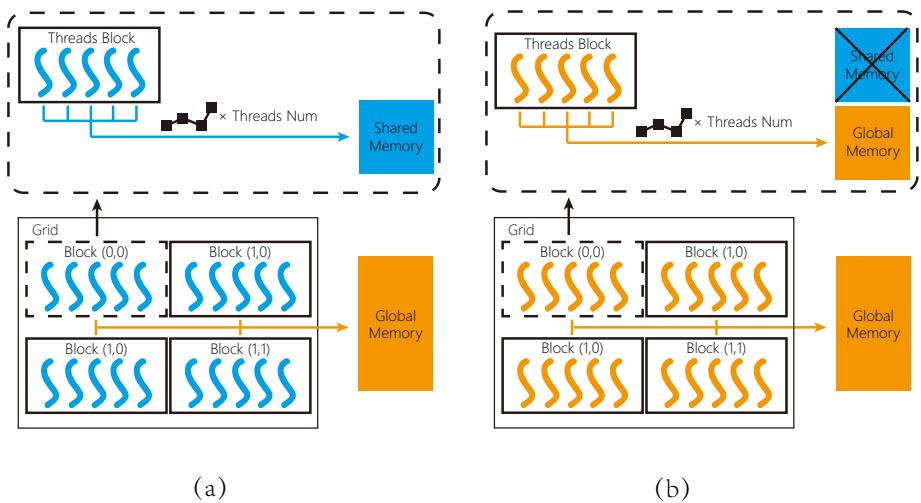}
	\caption{Memory structure and data flow between global memory, shared memory and threads of GPU.}
	\label{fig:memory}
\end{figure}

\subsubsection{Real space realization}

Realization of the method in real space is similar to the lattice realization, except the probability definition of the Rosenbluth weight, more precisely, the probability of positioning of a new monomer during chain generation. In real space the Rosenbluth weight is proportional to the volume available for the placement of a next monomer. 
The volume is estimated using Monte Carlo method: a bead is randomly placed $N_{trial}$ attempts at a fixed distance from the previous bead and the number $N_{allowed}$ (self-avoided) successful positions is counted. If $N_{allowed}>$ 0, a new position is accepted with the weight 1/$N_{allowed}$; the weight of the conformation $W_a$ is multiplied by the factor $1/N_{allowed}$. If there is no possibility to place a monomer, $N_{allowed}=0$, the generation restarts from the beginning.

\floatname{algorithm}{Algorithm}
\renewcommand{\algorithmicrequire}{\textbf{Require:}}
\renewcommand{\algorithmicensure}{\textbf{Ensure:}}

\begin{algorithm}
	\setstretch{1.00}
	\caption{Off-Lattice Rosenbluth chain generation}
	\begin{algorithmic}[1] 
		\Require $BlockDim.x$ is the block dimension defined by user
		\Require $seed$ is a random number seed chosen by the user
		\Require $chainlength$ is the polymer chain length chosen by the user
		
		\Require \textbf{Function} $rngonsphere$ generates a random position for the next monomer
		\Require \textbf{Function} $distance$ calculates the distance between two monomers
		\State $x,y,z \gets BlockDim.x * chainlength$
		\State $w \gets BlockDim.x$
		\State $rng \gets rng(seed)$
		\Function {Chaingeneration}{}
%
		
		\For{$i = 0 \to chainlength$}
		\If {$i == 1$}
		\State $Pos[0] \gets 0$ \Comment{$Pos[i]$ denotes position of monomer $i$ $(x_i,y_i,z_i)$}
		
		\State $p \gets 1$ \Comment{$p$ denotes Rosenbluth weight}
		\EndIf

		\For{$k = 0 \to ptMax$} \Comment{$ptMax$ denotes total trial attempts}
		\State $Pos[Temp] \gets rngonsphere$
		\State $Pos[Temp]=Pos[i-1]+Pos[Temp]$
		\State $overlap \gets \False$
		
		\For{$j = 0 \to i-1$}
		\State $R2 \gets distance(Pos[i],Pos[j])$
		\If {$R2 \leq d$} \Comment{$d$ denotes the polymer bondlength}
		\State $KK=KK+1$
		\State $overlap \gets \True$
		\EndIf
		\EndFor
		\If {$\neg overlap $}
		\State $Pos[i]=Pos[Temp]$
		\EndIf
		\EndFor
		\State $p=p*(1.0-kk/ptMax)$
		\EndFor

		\EndFunction
		
	\end{algorithmic}
\end{algorithm}

A block diagram illustrating a the generation process is shown in Figure \ref{fig:memory}a. Computation is performed by sets of streaming multiprocessors, each containing several computer units. Code is executed as a block of threads on a particular multiprocessor. Blocks  of threads are grouped in a grid. Each multiprocessor contains a small shared memory store that can be accessed by all threads in a given block. For current implementation we define the size of the block. The total number of blocks or the grid will automatically be defined as the total number of chains dividing the block dimension. On a single block, each thread simultaneously and synchronously generates its own statistically uncorrelated conformation of the molecule using the Rosenbluth algorithm. All the data are  stored in the shared memory during the generation process. The data in the shared memory can also be used on the fly: data is processed directly and then discarded. Alternatively, It can also be saved to the global memory on GPU and later copied back to CPU. However, since all coordinates are stored in shared memory during the SAW process, high consumption of shared memory may greatly affect the occupancy of the program.

\subsection{Performance}
In order to compare the performance of our GPU implementation, we implemented a sequential CPU version of the same algorithm. Similar to GPU realization, we use the single precision for the calculations. 

\begin{table}[h]
	\centering
	\caption{Comparison of performance between CPU (i5) and GPUs (K80/GTX 1080) for generation of 1 million linear polymer conformation in 3D lattice space}
	\label{latticeperf}
	\begin{tabular}{@{}llllll@{}}
		\toprule
		\multicolumn{1}{c}{\multirow{2}{*}{N} }& \multicolumn{3}{c}{Time for generation of $10^6$ chains (seconds)} & \multicolumn{2}{c}{\multirow{2}{*}{Acceleration}} \\ \cmidrule(lr){2-4}
		& \multicolumn{2}{c}{GPU}            & CPU           &                                \\ \cmidrule(r){1-6}
				Shared Memory Version  &K 80& GTX 1080&&K 80& GTX 1080
		\\\cmidrule(r){1-6}
		4                                       & 0.00676  &   0.00324       & 12.8489             & 1900.6    &3965.4                  \\
		8                                       & 0.01789   &   0.00901            & 31.2671              & 1747.7      &  3469.5              \\
		16                                      & 0.06670   &   0.02664             & 88.8572              & 1332.1      &   3335.0              \\
		32                                      & 0.41466   &   0.09192            & 192.644             & 464.6      &    2095.7             \\
		64                                      & 3.17465  &   0.38830            & 546.526            & 172.2      &     1407.5           \\
		128                                     &   8.32345    & 1.73263             & 1615.21      &194.6       & 932.2                     \\
		256                                     &   40.2882    & 7.50125            & 5285.14    &131.5         & 705.4                       \\
		512                                      &   225.873   & 49.8232            & 18570.6     &82.2        & 372.2                       \\
		1024                                    &   1921.74     & 373.824           & 70188.3      &36.5      & 188.1                       \\
		2048                                    &   11304.5     & 3981.55           & 267401.5     &23.7       & 67.2                       \\
		 \cmidrule(r){1-6}
		Global Memory Version  &K 80& GTX 1080&&K 80& GTX 1080
		\\\cmidrule(r){1-6}
		100                                      & 31.9       & 8.1    &  1140             & 36.8        & 140.7             \\	
		500                                      & 481.7     &  14.5     &  17400             & 36.3       &     119.4           \\	
		1000                                      & 1928.0     &  559.9     &  66200             & 34.3         &     118.2         \\	
		2500                                      & 12158.0    &  3574.0      &  395280             & 32.5         &    110.6          \\		
		5000                                      & 47775.3     &   14088.0    & 1552784             & 32.5             &   110.2       \\
		10000                                      & 189862.6   &   55883.1     &      6157540        & 32.4       &      110.2          \\ \bottomrule
	\end{tabular}
\end{table}

\begin{table}[h]
	\centering
	\caption{Comparison of performance between CPU (i5) and GPUs (K80/GTX 1080) for generation of 1 million linear polymer conformation in real space}
	\label{my-ofl}
	\begin{tabular}{@{}llllll@{}}
		\toprule
\multicolumn{1}{c}{\multirow{2}{*}{N} }& \multicolumn{3}{c}{Time for generation of $10^6$ chains (seconds)} & \multicolumn{2}{c}{\multirow{2}{*}{Acceleration}} \\ \cmidrule(lr){2-4}
& \multicolumn{2}{c}{GPU}            & CPU           &                                \\ \cmidrule(r){1-6}
Shared Memory Version (GPU: K80) &K80& GTX 1080&&K80& GTX 1080
\\\cmidrule(r){1-6}
		4                                       & 0.20985   & 0.16288         & 36.023              & 171.7            & 221.2           \\
		8                                       & 0.5912      & 0.44877        & 94.344              & 159.6            & 210.2           \\
		16                                      & 2.98727     & 2.31603        & 259.478             & 86.9           & 112.0            \\
		32                                      & 26.19131     & 11.4375       & 655.682             & 25.0            & 57.4            \\
		64                                      & 315.1483    & 58.3319        & 1967.91            & 6.2& 33.7 \\
		128                                      & 1255.879 & 226.441             & 6288.21         & 5.0     & 27.8                     \\
		256                                      & 6099.337 & 986.501            & 22638.1        & 3.7      & 23.0                       \\
		512                                      & 39547.49 & 5905.82            & 83398.6     & 2.1         & 14.1                      \\
		1024                                     & 114069.2  & 45513.8           & 305410     & 2.7        & 6.7                      \\
		2048                                     & 1232172.5  & 378758           & 1199200   & 1.0          & 3.2                       \\
        \cmidrule(r){1-6}
       Global Memory Version
       \\\cmidrule(r){1-6}
		100                                      & 1994        & 346     &  3990             & 2.001           & 11.5           \\		
		500                                      & 45261     & 8918        &  75810             & 1.674           & 8.5            \\		
		1000                                      & 175970     & 35277        &  292800             & 1.663       & 8.3                \\		
		2500                                      & 1125674  & 216963           &  1779100             & 1.580         & 8.2              \\		
		5000                                      & 4464250   & 860729          & 7057980             & 1.581        & 8.2                \\
		10000                                      & 14336106    & 3629576         &      29762520        & 1.313        & 8.2              \\ \bottomrule 
	\end{tabular}
\end{table}

A direct comparison between CPU and GPU of both lattice and off-lattice version is presented in Tables \ref{latticeperf} and \ref{my-ofl}. We define a speed-up factor as follows: tCPU is the execution time on a single CPU core and tGPU is the runtime on the GPU. During the run one million of linear polymer conformations of different lengths is generated on NVIDIA Tesla K80 GPU or NVIDIA GTX 1080 and Intel i5-3320m CPU. With growing length, the efficiency gradually decreases. The decrease comes mainly from the increasing consumption of the shared memory of each block. However, even for long polymers with 2048 monomers GPU outperform single core CPU as a factor of 67.2.

The real space realization is less efficient because MC process of calculating Rosenbluth weights leads to assynchronization of the code (for example, "if" statements).

\begin{figure}[H]
	\centering
	\includegraphics[width=16cm]{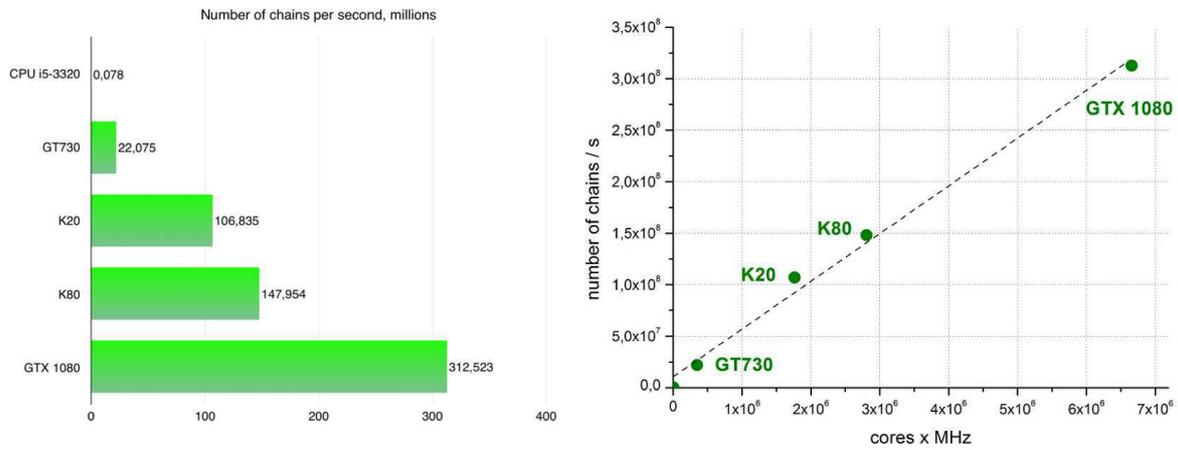}
	\caption{Comparison of the average number of chains generated per second between different architectures for linear chains comprised 4 monomers on a lattice. The hardware details is presented in Table \ref{tab:expcond} }
	\label{fig:perf}
\end{figure}

Benchmarks on different GPUs are summarized in Table \ref{tab:expcond} and the comparison of performance is presented in Figure \ref{fig:perf}. The performance is evaluated as the number of chains generated per second.

\begin{table}[htbp]
	\centering
	\caption{Hardware configurations for benchmarks test}
	\label{tab:expcond}
	\begin{tabular}{ccccc}
		\toprule
		& GTX 1080 &K80 &  K20  & GT 730 \\
		\midrule
		GPU &GTX 1080& K80 & K20& GT 730 \\
		Stream Processors &3840& 2 x 2496 & 2496 & 384\\
		Core clock &1734 MHz& 562 MHz & 706 MHz & 902 MHz\\
		Memory clock &5 GHz& 5 GHz & 5.2 GHz & 5.0 GHz\\
		DRAM &8 GB GDDR5& 2 x 12 GB GDDR5 & 5 GB GDDR5 & 1 GB GDDR5\\
		Shared Memory &96 KB& 128 KB & 64 KB & 32 KB\\
		\bottomrule
	\end{tabular}
\end{table}

One can see from Tables \ref{latticeperf},\ref{my-ofl} that for shared memory method implementation, the increasing chain length leads to increase of consumption of shared memory. Thus, the number of blocks running simultaneously will be reduced to meet the increasing  shared memory consumption since the total shared memory size for each GPU is fixed.

For relatively long chains, we propose data flow shown in Figure \ref{fig:memory}b where all coordinates are stored in global memory instead of shared memory. Although the data transfer between global memory is much slower than shared memory, global memory is larger thus can accomodate larger molecules. A detailed comparison of the efficiency between global memory implementation and CPU implementation is shown in Tables \ref{latticeperf},\ref{my-ofl}. 
In such case, we find that although the bandwidth of global memory is still quite slow compared with shared memory, it will not affect the number of blocks running simultaneously. As a result, the global memory implementation will be faster for long chains that do not fit in shared memory.

With new NVIDIA Pascal architecture, the shared memory is increased due to the larger number of Stream Multiprocessor (SM) count, and aggregate shared memory bandwidth is effectively more than doubled. A higher ratio of shared memory, registers, and warps per SM is introduced to the new benchmark GP100 which allows the SM to more efficiently execute code. The bandwidth per-thread to shared memory is also increased. All these features will increase the performance of our implementation.

\section{Practical example}
\subsection{Static properties of polymer melts}
As an example of accuracy and efficiency of the method, we investigated classical properties of polymer melts. The averaged squared extension, $R^{2}$, of the chain is calculated for all the chains generated and compared as a function of $N$. $R^{2}$ is also known as end-to-end distance \cite{flory_principles_1953} which is of importance in calculating the properties such as viscosity of the polymer chains. For a diluted solution of polymer chains the dependence of this quantity on the number of monomers in the chain is given by \cite{doi_theory_1988}:
\begin{equation} 
	\label{eu_eqn}
	\big \langle R_{N}^2 \big \rangle = a N^{2\nu}
\end{equation}
where the proportionality constant $a$ depends on the structure and on external conditions such as the solvent used in the chemical solution or temperature; critical exponent $\nu$ is universal, and depends only on the dimension of space.

\begin{figure}[H]
	\centering
	\includegraphics[width=12cm]{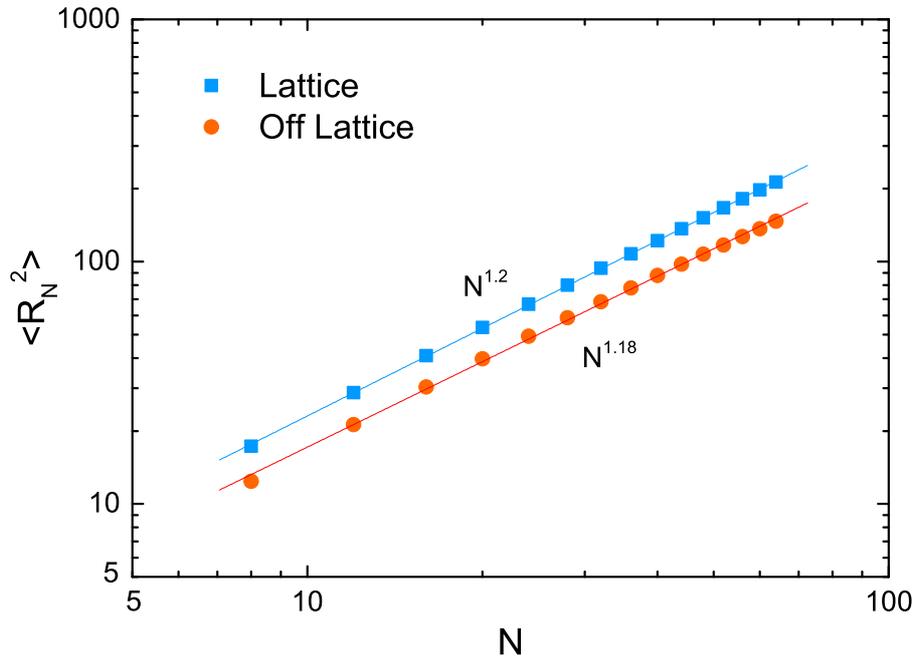}
	\caption{Average end-to-end distance $R^2$ of a polymer chain as a function of polymer chain length $N$.}
	\label{fig:ndis}
\end{figure}

We calculated $R^{2}$ in the case of both lattice and off-lattice versions with various values of $N$ up to 64 in three dimensions. Results are shown in Figure \ref{fig:ndis}, and compared with previous studies. We find that the critical exponent $\nu$ for lattice model is 0.601, and 0.588 for off-lattice model. The above results give a remarkably good match in entire range with  Ref. \citenum{gould_introduction_2007}.

We also performed series of simulation to test our code to calculate the partition function of the linear polymers $Z_N$ with scaling arguments of SAW. A general superscaling expression for partition function of the system of length $N$ is given by\cite{duplantier_statistical_1989}

\begin{equation}
Z_{N}\sim\mu ^{N}N^{\gamma-1}
\label{eq:scale}
\end{equation}
where $\mu$ is a model dependent connectivity constant and $\gamma$ is the universal entropic exponent. Consider a sample of $\aleph$ configurations of polymers with length $N$, ($s_1,s_2,...,s_\aleph$) and corresponding Rosenbluth weights $(W(s_i))$. Then $Z_N$ is estimated as

\begin{equation}
Z_{N}\approx \left\langle W_{N}\right\rangle _{\aleph}=\dfrac {1} {\aleph}\sum _{i=1}^{\aleph}W\left( s_{i}\right) 
\end{equation}

Taking logarithm of both sides of eq \ref{eq:scale} we can get an estimate for exponent $\gamma -1$ and constant $\mu$:

\begin{equation}
\log\left( Z_N \right) \sim N\log\left( \mu \right) +\left( \gamma -1\right) \log{\left( N\right) }
\end{equation}

 We performed series of simulation in 3D lattice space and found a best estimate $\mu =4.6856$ and $\gamma = 1.1592$ which is in a good agreement with previous studies \cite{grassberger_monte_1993,hsu_scaling_2004,clisby_self-avoiding_2007} . The results are summarized in Table \ref{my-label}.

\begin{table}[]
	\centering
	\caption{Comparison of SAW exponents in 3D lattice space between values from literature and obtained from static MC simulation performed on GPU }
	\label{my-label}
	\begin{tabular}{ccc}
		\toprule
		& Literature values & Static MC Simulation \\
		\midrule
		$\gamma$         & 1.1608 \cite{grassberger_monte_1993}          &    $1.1592 \pm 0.0006$    \\
		$\nu$ & 0.5877 \cite{hsu_scaling_2004}          &   $0.5880  \pm  0.0001$     \\
		$\mu$            & 4.684 \cite{clisby_self-avoiding_2007}          &    $4.6856 \pm  0.0001$ \\
		\bottomrule     
	\end{tabular}
\end{table}

\section{Conclusion}

We developed highly efficient parallel GPU implementation of the Rosenbluth algorithm of generation of self-avoiding random walks on a lattice and in real space, which scales almost linear with the number of CUDA cores. Both versions of the code have the same accuracy compared with a single core CPU implementation, but give huge performance improvement in simulation efficiency making the generation process almost perfectly parallelizable \cite{guo_computer_2016}. The implementation breaks the performance bottleneck of existing molecular conformation generating methods, significantly improving  parallel performance, and has broad application prospects in the static MC simulations. In molecular simulations, the whole configurational sampling of short lipids, surfactants or peptide sequences can be generated directly. 

The performance is larger for smaller system sizes due to limitations of size of shared memory and number of registers in existing GPU architectures. With growing computational capacities of GPUs one can expect further increase of efficiency allowing for more and more practical applications of the presented method. Thus, this method could potentially find applications in many other fields such as robotics, artificial intelligence, big data analysis.  

\section*{Acknowledgements}
Authors acknowledge funding from Marie Curie Actions under EU FP7 Initial Training Network SNAL 608184 and 
High Performance Computing center of the University of Strasbourg for access to computing resources. Authors are grateful to Dr. Olivier Benzerara for useful discussions and  Dr. Marco Werner for help in presentation of results in Fig. 3.

\section*{Suplementary information}

\begin{itemize}
  \item Suppinfo.pdf: Parallel implementation of PERM on GPU.
\end{itemize}


\end{document}